\documentclass[a4paper,11pt,dvips]{article}
\usepackage{graphicx}
\usepackage{amsmath}
\usepackage{amssymb}
\usepackage{latexsym}
\usepackage{color}

\newcommand{\bra}{\begin{array}}
\newcommand{\era}{\end{array}}
\newcommand{\beq}{\begin{equation}}
\newcommand{\eeq}{\end{equation}}
\newcommand{\beqar}{\begin{eqnarray}}
\newcommand{\eeqar}{\end{eqnarray}}

\def\BC{\bb C}
\def\_\BC{\bbi C}

\def\Tr {{\rm Tr}}

\def\( {\left(}
   \def\) {\right)}
\def\[ {\left[}
\def\] {\right]}
\def\no2 {{\textstyle{n\over 2}}}

\def\Tr {{\rm Tr}}
\def\dag {{\dagger}}

\newcommand{\om}{\omega}

\newcommand{\lam}{\lambda}

\newcommand{\be}{\beta}

\newcommand{\te}{\theta}

\newcommand{\al}{\alpha}

\newcommand{\da}{\dagger}

\newcommand{\ov}{\over}
\newcommand{\hb}{\hbar}

\newcommand{\ev}{\equiv}
\newcommand{\lb}{\label}

\begin{document}
\begin{titlepage}
\setcounter{page}{1}
\renewcommand{\thefootnote}{\fnsymbol{footnote}}

\begin{center}

\begin{flushright}
ucd-tpg:1106.06\\
\end{flushright}

\vspace{0.5cm}
 {\Large \bf 
 Entanglement in Coupled  Harmonic Oscillators \\ Studied Using  a  Unitary Transformation} 

\vspace{0.5cm}

{\bf Ahmed Jellal$^{a,b,c}$\footnote{{\sf ajellal@ictp.it -  jellal.a@ucd.ac.ma}}},
{\bf Fethi Madouri$^{a,d}$} and {\bf Abdeldjalil Merdaci$^{a,e}$}

\vspace{0.5cm}

$^a${\em {\it Physics Department, College of Science, King Faisal University,\\
PO Box 380, Alahsa 31982,
Saudi Arabia}}

$^b${\em Saudi Center for Theoretical Physics, Dhahran, Saudi Arabia}

$^{c}${\em Theoretical Physics Group,  
Faculty of Sciences, Choua\"ib Doukkali University,\\
PO Box 20, 24000 El Jadida,
Morocco} 

$^{d}${\em IPEIT, Rue J. Lal Nehru, Montfleury, University of Tunis, Tunisia}

$^{e}${\em  D\'epartement des Sciences Fondamentales, 
 Universit\'e 20.08.1955 Skikda,\\
  BP 26, DZ-21000 Skikda, Algeria}

\vspace{3cm}

\begin{abstract}

We develop an approach  to study
the entanglement in two coupled  harmonic oscillators.
We start by introducing
an unitary transformation
to end up with the solutions of the energy spectrum. These are used
to construct the
corresponding coherent states through the standard way.
To evaluate the degree of the entanglement between the obtained states,
we calculate the purity function in terms of the coherent
and number states, separately. The result is yielded two parameters
dependance of the purity, which can be controlled easily. Interesting
results are derived by fixing the mixing angle of such transformation
as $\frac{\pi}{2}$.
We compare our results with already published work and point out the relevance
of these findings to a systematic formulation of the entanglement effect in two
coupled harmonic oscillators.

\end{abstract}
\end{center}

\vspace{3cm}

\noindent PACS numbers: 03.65.Ud, 03.65.-w, 03.67.-a

\noindent Keywords: harmonic oscillator, unitary transformation, purity, entanglement.

\end{titlepage}

\newpage
\section{Introduction}

Entanglement is one of the most remarkable features of quantum mechanics that does not have any classical counterpart.
It is a notion which has been initially introduced and coined by Schr\"odinger \cite{1} when quantum mechanics was still
in its early stage of development. Its status has evolved throughout the decades and has been subjected to significant
changes. Traditionally, entanglement  has been related to the most quantum mechanical exotic concepts such as
Schr\"odinger cat \cite{1},
Einstein-Podolsky-Rosen paradox \cite{2} and violation of Bell's inequalities \cite{3}.
Despite its conventional significance, entanglement has gained, in the last decades, a renewed interest mainly  because
 of the development of the quantum information science \cite{4}. It has been revealed that it lies at the heart of various
 communication and computational tasks that cannot be implemented classically. It is believed that entanglement  is the main
 ingredient of the quantum speed-up in quantum computation \cite{4}. Moreover, several quantum protocols such as teleportation,
 quantum dense coding, and so on \cite{Ben1,Ben2,Eckert,Murao,Fuchs,Rausschendorf,Gottesman}  are exclusively realized with the
 help of entangled states.
With this respect, many interesting works appeared
dealing with
the development of
a quantitative theory of entanglement and the definition of its
basic measure. These concern the concurrence, entanglement of formation and linear
entropy \cite{Rungta,Ben3,Wootters,Coffman}.

Entangled quantum
systems can exhibit correlations that cannot be explained on the
basis of classical laws and the entanglement in a collection of
states is clearly a signature of non-classicality \cite{Markham}.
Furthermore, in the last few years it has become evident that quantum information may lead to further
insights into other areas of physics \cite{18}. This has led to a cross-fertilizing between different
areas of physics. It is worthy of note that the nonlinear Kerr effect
\cite{19} has been considered as the most famous source of physical realization of photon pairs of
entangled polarization states. However, it raises a number of difficulties to the control of photons
that are traveling at the speed of light. This is why so much attention has been paid recently to the
entangled states of massive particles as they are viewed to be much more easy to control 
\cite{18,harshman}.

On the other hand,
the harmonic oscillator machinery plays a crucial role in many areas of physics. These are
the Lee model in quantum field
theory~\cite{sss61}, the Bogoliubov transformation in
superconductivity~\cite{fewa71}, two-mode squeezed states of
light~\cite{hkn90,dir63,cav85}, the covariant harmonic oscillator
model for the parton picture~\cite{kim89}, and models in molecular
physics~\cite{iac91}. {There are also models 
in which one of the variables is not observed, including
thermo-field dynamics~\cite{ume82}, two-mode squeezed
states~\cite{yupo87,ekn89}, the hadronic temperature~\cite{hkn89},
and the Barnet-Phoenix version of information
theory~\cite{baph91}. These physical models are the examples of
Feynman's rest of the universe. In the case of two coupled harmonic oscillator,
the first one is the universe and the second one is the rest of universe.
For sake of the mathematical simplicity, the mixing angle (rotation of the coordinate system), in the above mentioned references,
is taken to be equal   ${\pi\ov 2}$. This means that the system consists of
two identical oscillator coupled together by a potential term.}



In the context of the entangled massive particles, 
we cite the recently achieved investigation of
a specific realization of {two coupled} harmonic oscillator model by the authors of reference \cite{harshman}.
In fact, they calculated the interatomic entanglement for Gaussian and non-Gaussian pure states 
by using  the purity function
of the reduced density matrix. This allowed them to treat the cases of free and trapped molecules and hetero- and homonuclear
molecules.
Finally, they concluded that when the trap frequency and the molecular frequency are very different,
and when the atomic masses are equal, the atoms are highly-entangled for molecular coherent states and number states.
Surprisingly, while the interatomic entanglement can be quite large even for molecular coherent states,
the covariance of atomic position and momentum observables can be entirely explained by a classical model
with appropriately chosen statistical uncertainty.

Motivated by the mentioned references above and in particular \cite{harshman}, we undertake to develop a new approach to
study the entanglement in two coupled harmonic oscillators. It is based on a suitable transformation having the merit
of reducing the relevant physical parameters into two: the coupling parameter
$\eta$ and  mixing angle  $\te$. It turns out that we can easily derive the solutions corresponding to the energy spectrum.
Then, the obtained solutions are used to construct the coherent states through the standard method.  In order to
characterize the degree of entanglement, we calculate, within the framework of the coherent states, the purity function. Then the final
form of the purity is cast {in terms of $\eta$  and  $\te$}.
Our finding shows two interesting results:
the first one tells us that the present system is not entangled at
$\eta=0$, as expected, and highly entangled at large $\eta$ (Figure 1).
The second one  is when we fix $\te=\frac{\pi}{2}$, the purity behaves like the inverse of
$\cosh\eta$ and the corresponding plot (Figure 2) shows that the purity is ranging between  0 and  1. It is worthy of notice that,
in this case, the purity becomes one parameter dependent, which means that it is easy to control.


Subsequently, we evaluate the purity in terms of the number states. In doing so, we
use the well-known relation to express the number states $|n_1,n_2\rangle$ as function of the corresponding
coherent states $|\al,\be\rangle$. 
Then after a lengthy but a straightforward algebra,
we end up with the final
form of purity. To be much more concrete, we restrict ourselves to some interesting cases
that are $(n_1=1,n_2=0)$ and $(n_1=1,n_2=1)$. For the first configuration,
the obtained purity is simply a ratio of hyperbolic and sinusoidal functions,
which  tells us that the entanglement is maximal at large $\eta$ for all $\te$
(Figure 3).
 In the particular
case when $\te=\frac{\pi}{2}$, the purity is typically a ratio of a hyperbolic cosine  function,
which  shows clearly that the purity is positive as it should be (Figure 4).
The second configuration gives also a mixing dependence between the hyperbolic and sinusoidal functions
where the corresponding plots (Figures 5 and 6) show some difference in the form with respect to the
first one. In both cases, we notice that the numerators are always
 hyperbolic  cosine of even $\eta$ and denominators are also power of
 the function $\cosh{\eta}$.

The present paper is organized as follows. In section 2, we review the derivation
of the solutions of the energy spectrum of two coupled harmonic oscillators~\cite{jellal}.
These will be used to build the corresponding coherent states and therefore
evaluate the purity function of the reduced matrix elements in section 3.
The final form of purity function is subjected to different investigations where
we underline its dependence to two physical parameters $\eta$ and $\te$. 
In section 4, we evaluate the purity in terms of the number states after a series
of transformation. Two interesting case of the purity will be discussed in section 5.
Finally, we give conclusion and perspective of our work.

\section{Energy spectrum solutions} 

In doing our task, we
consider a system of two coupled harmonic oscillators
parameterized by the planar coordinates $(X_1, X_2)$ and masses
$(m_1,m_2)$. Accordingly, the corresponding Hamiltonian is written
as the sum of free and interacting parts
\cite{hkn99am}
\begin{equation}\label{HAM1}
H_1 = {1\over 2m_{1}}P^{2}_{1} + {1\over 2 m_{2}} P^{2}_{2} +
{1\over 2} \left( C_1X^{2}_{1} + C_2 X^{2}_{2} + C_3 X_{1}
X_{2}\right)
\end{equation}
where $C_1, C_2$ and $C_3$ are constant parameters.
After rescaling the position variables
\begin{equation}
x_{1} = \mu X_{1},
\qquad
x_{2} = \mu^{-1} X_{2}
\end{equation}
as well as the momenta
\begin{equation}
p_{1} = \mu^{-1} P_{1},
\qquad
p_{2} = \mu P_{2}
\end{equation}
$H_1$ can be written as
\begin{equation}\label{HAM2}
H_2 = {1\over 2m}\left(p^{2}_{1} + p^{2}_{2} \right) + {1\over
2}\left( c_1 x_{1}^{2} + c_2 x^{2}_{2} + c_3 x_{1} x_{2} \right)
\end{equation}
where the parameters are given by
\begin{equation}\label{para}
\mu=({m_{1}/ m_{2}})^{1/ 4}, \qquad m = (m_{1}m_{2})^{1/2},\qquad c_1=C_1\sqrt{m_2\over m_1}, \qquad
c_2=C_2\sqrt{m_1\over m_2},\qquad c_3=C_3.
\end{equation}

As the Hamiltonian~(\ref{HAM2}) involves an interacting
term, a straightforward investigation of the basic features of the
system is not easy. Nevertheless, we can simplify this
situation by a transformation to new phase space variables
\begin{equation}
\lb{TRAN} y_a = M_{ab} x_b, \qquad  \hat{p}_a = M_{ab} p_b
\end{equation}
where the matrix
\begin{equation} \label{UMAT}
(M_{ab}) =
\left(%
\begin{array}{cc}
\cos {\te\ov 2} & -\sin {\te\ov 2} \cr
\sin {\te\ov 2} & \cos {\te\ov 2}\cr
\end{array}
\right).
\end{equation}
is a unitary rotation with the mixing angle
$\te$. Inserting the mapping~(\ref{TRAN}) into~(\ref{HAM2}), one
realizes that $\te$ should satisfy the condition
\begin{equation}\label{acon}
\tan \te  = {c_3\over c_2 - c_1}
\end{equation}
to get a factorizing Hamiltonian
\begin{equation}
\label{HAM3} H_3 = {1\over 2m} \left(\hat{p}^{2}_{1} +
\hat{p}^{2}_{2} \right) +
{k\over 2}\left(e^{2\eta } y^{2}_{1} + e^{-2\eta }
y^{2}_{2}\right)
\end{equation}
where we have introduced two parameters
\begin{equation}\label{PARA}
k = \sqrt{c_1c_2 - c_3^{2}/4} , \qquad  e^{2\eta}= \frac {c_1 + c_2
+ \sqrt{(c_1 - c_2)^{2} + c_3^{2}}}{2k}
\end{equation}
under the reserve  that the condition $4c_1c_2 > c_3^{2}$ must be fulfilled.
{The parameter $\eta$ is actually measuring the strength of the coupling.}

For later use, it is convenient to
separate the Hamiltonian~(\ref{HAM3}) into two commuting parts
and then write $H_3$ as
\begin{equation}
\label{HAM4} H_3 = e^{\eta} {\cal H}_1 + e^{-\eta} {\cal H}_2
\end{equation}
where ${\cal H}_1$ and ${\cal H}_2$ are given by
\begin{equation}
\label{HAM41} {\cal H}_1 = {1\ov 2m}e^{-\eta}\hat{p}^{2}_{1} +
{k\ov 2} e^{\eta}y^{2}_{1} , \qquad
{\cal H}_2 = {1\ov 2m}e^{\eta}\hat{p}^{2}_{2} +
{k\ov 2} e^{-\eta}y^{2}_{2}.
\end{equation}
One can see that the decoupled Hamiltonian
\begin{equation}
\label{HAM5} H_0 = {1\ov 2m}\hat{p}^{2}_{1} +
{k\ov 2} y^{2}_{1} + {1\ov 2m}\hat{p}^{2}_{2} +
{k\ov 2}y^{2}_{2}
\end{equation}
is obtained for $\eta=0$, which is equivalent to set $c_3=0$.

The Hamiltonian $H_3$ can simply be diagonalized
by defining a set of  annihilation and creation operators. These are
\begin{equation}
\lb{CRAN}
a_i = \sqrt{k\ov 2\hb\om} e^{\varepsilon\eta\ov 2}y_{i} +
{i \ov \sqrt{2m\hb\om}} e^{-{\varepsilon\eta\ov 2}}\hat{p}_{i},
\qquad a_i^{\da} = \sqrt{k\ov 2\hb\om} e^{\varepsilon\eta\ov 2}y_{i}
- {i \ov \sqrt{2m\hb\om}}
e^{-{\varepsilon\eta\ov 2}}\hat{p}_{i}
\end{equation}
with {the} frequency
\begin{equation}
\om=\sqrt{k\ov m}
\end{equation}
{and} $\varepsilon=\pm 1$ for $i=1,2$, respectively.
They satisfy the commutation relations
\begin{equation}
 [a_i, a_j^{\dag}] = \delta_{ij}
\end{equation}
whereas other commutators vanish. Now we can map $H_3$ in terms of
$a_i$ and $a_i^{\dag}$ as
\begin{equation}
\label{HAM7} H_3= \hb\om \left( e^{\eta }a_1^{\dag}a_1 + e^{-\eta}a_2^{\dag}a_2 +
\cosh\eta\right).
\end{equation}

To obtain the eigenstates and the eigenvalues, one
solves the eigenequation
\begin{equation}
 H_3|n_1, n_2\rangle = {\cal E}_{n_1,n_2} |n_1,
n_2\rangle
\end{equation}
getting the states
\begin{equation}\label{frst}
|n_1,n_2\rangle= {(a_1^{\dag})^{n_1} (a_2^{\dag})^{n_2} \ov
\sqrt{n_1!n_2!}} |0, 0\rangle
\end{equation}
as well as the energy spectrum
\begin{equation}
\label{SPE2} E_{3,n_1,n_2} = {\hb\om} \left(e^{\eta}n_1  +e^{-\eta} n_2 +\cosh\eta\right).
\end{equation}
It is clear that these eigenvalues reduce to {those of} the decoupled harmonic oscillators,
namely ${\hb\om} \left(n_1  +  n_2 +1\right)$. This shows clearly that the presence
of the coupling parameter $\eta$ will make difference and allow us to
derive interesting results in the forthcoming
analysis.

To show the correlation between variables, let us just focus on the ground state
and write the corresponding wavefunction in $y$-representation.
This is
\begin{equation}
\label{YWAV0}
\psi_0(\vec{y})\ev \langle y_1,y_2|0,0\rangle
= \sqrt{m\om \ov \pi\hb}
\exp{\left\{-{m\om \over 2\hb}\left(e^{\eta} y^{2}_{1} + e^{-\eta}
y^{2}_{2}\right) \right\} }
\end{equation}
which  can easily be used to deduce
the ground state wavefunction in terms of
the variables $(x_1,x_2)$. Therefore, from the unitary
representation we find 
\begin{eqnarray}
\lefteqn{
\lb{XWAV} \psi_{0}(\vec{x}) \ev
\langle x_1,x_2|0,0\rangle}
\nonumber \\
& &
~~
= \sqrt{m\om \ov \pi\hb}
\exp\left\{-{m\om \over 2\hb}\left[e^{\eta}\left(x_{1}
\cos{\te\over 2} - x_{2} \sin{\te\over 2}\right)^{2} +
e^{-\eta}\left(x_{1}\sin{\te\over 2} + x_{2}\cos{\te\over
2}\right)^{2} \right] \right\}.
\end{eqnarray}
We notice that~(\ref{YWAV0}) is separable in terms of the variables
$y_{1}$ and $y_{2}$, which is not the case
for~(\ref{XWAV}) in terms of $x_{1}$ and $x_{2}$.
We close this part by claiming that
the  obtained results so far will be used
to study the entanglement in the present system.

\section{ Entanglement in coherent states}

As we claimed above, 
{we implement} our approach to study the entanglement
of two coupled harmonic oscillators. Actually, it can be seen as another alternative method to
recover the results obtained in~\cite{harshman} not only in a simpler way but also with less physical
parameters of control.
To start let us first introduce the coherent states corresponding to
the eigenstates $|n_1,n_2\rangle$ given in~(\ref{frst}). As usual,
we can use
 the displacement operator to define the coherent states in terms of two complex
 numbers $\al$ and $\be$. These are
\beq
|\al,\be\rangle= D(a_1,\al) D(a_2,\be) |0,0\rangle
\eeq
which gives the wavefunction
\beqar\lb{wfy}
\Phi_{\alpha\beta}\left(  y_{1},y_{2}\right)    =\left(  \frac{\lam
_{1}\lam_{2}}{\pi}\right)  ^{1/2}\exp\left[  -\frac
{\lam_{1}^{2}}{2}y_{1}^{2}-\frac{\left\vert \alpha\right\vert ^{2}}%
{2}- \frac{\alpha^2}{{2}}+\sqrt{2}\al \lam_{1}y_{1}\right. 
\left.  -\frac{\lam_{2}^{2}}{2}y_{2}^{2}-\frac{\left\vert \beta
\right\vert ^{2}}{2}-\frac{\beta^2}{2}+ \sqrt{2}\beta  \lam_{2}y_{2}\right]
\eeqar
where we have set the quantities
\beq
\lam_{1}=e^{\frac{\eta}{2}}\left(  \frac{mk}{\hbar^{2}}\right)^{1/4}, \qquad
\lam_{2}=e^{-\frac{\eta}{2}}\left(  \frac{mk}{\hbar^{2}}\right)^{1/4}.
\eeq
In terms of the original variables $(X_1,X_2)$,  (\ref{wfy}) reads as
\beqar\lb{wf11}
\Phi_{\alpha\beta}\left(  X_{1},X_{2}\right)
 &=\left(  \frac{\lam
_{1}\lam_{2}}{\pi}\right)  ^{1/2}\exp\left[  -\frac
{\lam_{1}^{2}}{2}\left(  \mu\cos  \frac{\theta}{2}
X_{1}-\frac{1}{\mu}\sin  \frac{\theta}{2}  X_{2}\right)
^{2}-\frac{\lam_{2}^{2}}{2}\left(  \mu\sin  \frac{\theta}{2}
X_{1}+\frac{1}{\mu}\cos  \frac{\theta}{2}  X_{2}\right)
^{2}\right]  \nonumber\\
& \times \exp\left[\sqrt{2}\al  \lam_{1}\left(
\mu\cos  \frac{\theta}{2}  X_{1}-\frac{1}{\mu}\sin
\frac{\theta}{2}  X_{2}\right)  +\sqrt{2}%
\be  \lam_{2}\left(  \mu\sin  \frac{\theta}{2}
X_{1}+\frac{1}{\mu}\cos  \frac{\theta}{2}  X_{2}\right)\right] \nonumber\\
&\times\exp\left[-\frac{\left\vert \alpha\right\vert ^{2}}{2}-\frac{\left\vert \beta
\right\vert ^{2}}{2}-\frac{\alpha^2}{2}-
\frac{\beta^2}{2} \right].~~~~~~~~~~~~~~~~~~~~~~~~~~~~~~~~~~~~~~~~~~~~~~~~~~~~~~~~~
\eeqar
As it is clearly shown in the wavefunction (\ref{wf11}), the non-separability of
the variables will play in crucial role in discussing
the entanglement in the present system. This statement will be clarified later on when we will come
to the analysis of the role of the involved parameters.

At this level we have set all ingredients to study the
entanglement in the present system. All we need is to
determine explicitly the purity function that is a trace
of the density square corresponding to the obtained
eigenstates. More precisely, we have
\beq
P= \Tr{\rho}^2
\eeq
which in terms of the above coherent states reads as
\beq
P_{\alpha\beta}= \int dX_{1}dX_{1}^{\prime}dX_{2}dX_{2}^{\prime} \Phi_{\alpha\beta}\left(  X_{1},X_{2}\right)
\Phi_{\alpha\beta}^*\left(  X_{1}^{\prime},X_{2}\right) \Phi_{\alpha\beta}\left(  X_{1}^{\prime},X_{2}^{\prime}\right)
\Phi_{\alpha\beta}^*\left(  X_{1},X_{2}^{\prime}\right).
\eeq
Upon substitution, we obtain the form
\beqar
P_{\alpha\beta}   &=&\left(\frac{\lam_{1}\lam_{2}}{\pi}\right)^2 \int
dX_{1}dX_{1}^{\prime}dX_{2}dX_{2}^{\prime} \ \
 e^{-\mu^{2}\left(  \lam_{1}%
^{2}\cos^{2} \frac{\theta}{2}  +\lam_{2}^{2}\sin^{2}
\frac{\theta}{2}\right)    \left(X_{1}^{2}+X_{1}^{\prime2}\right)-\frac{1}{\mu^{2}}\left(
\lam_{1}^{2}\sin^{2}  \frac{\theta}{2}  +\lam_{2}^{2}%
\cos^{2}  \frac{\theta}{2}\right)    \left(X_{2}^{2}+X_{2}^{\prime2}\right)}\nonumber\\
&& \times e^{\frac{1}{2}\left(  \lam_{1}^{2}-\lam_{2}^{2}\right)
\sin\theta \left( X_{1}^{\prime}X_{2}+X_{1}X_{2}+X_{1}^{\prime}X_{2}^{\prime}+ X_{1}X_{2}^{\prime}\right)}
 \ \  e^{2\mu\left(  \frac{\alpha+\alpha^{\ast}}{\sqrt{2}}\lam_{1}%
\cos  \frac{\theta}{2}  +\frac{\beta+\beta^{\ast}}{\sqrt{2}%
}\lam_{2}\sin  \frac{\theta}{2}\right)    \left(X_{1}+
X_{1}^{\prime}\right)} \\
&& \times e^{-\frac{2}{\mu}\left(  \frac{\alpha+\alpha^{\ast}}{\sqrt{2}%
}\lam_{1}\sin  \frac{\theta}{2}  -\frac{\beta+\beta^{\ast}%
}{\sqrt{2}}\lam_{2}\cos \frac{\theta}{2}  \right) \left(X_{2}+
X_{2}^{\prime}\right)} \ \  e^{-2\left\vert \alpha\right\vert ^{2}-2\left\vert \beta\right\vert
^{2}-\alpha^{2}-\alpha^{\ast2}-\beta^{2}-\beta^{\ast2}}. \nonumber
\eeqar
This integral can easily be evaluated by introducing an appropriate transformation. 
 {This can be done by making use of} the following change of variables
\beq\lb{chav}
\left(
\begin{array}
[c]{c}%
X_{1}\\
X_{1}^{\prime}\\
X_{2}\\
X_{2}^{\prime}%
\end{array}
\right)  =\frac{1}{2}\left(
\begin{array}
[c]{cccc}%
\frac{\omega_{1}}{\mu\sqrt{1-2a}} & \frac{\sqrt{2}}{\mu}\omega_{1} &
\frac{\omega_{1}}{\mu\sqrt{1+2a}} & 0\\
\frac{\omega_{1}}{\mu\sqrt{1-2a}} & -\frac{\sqrt{2}}{\mu}\omega_{1} &
\frac{\omega_{1}}{\mu\sqrt{1+2a}} & 0\\
-\frac{\mu\omega_{2}}{\sqrt{1-2a}} & 0 & \frac{\mu\omega_{2}}{\sqrt{1+2a}} &
{\sqrt{2}}\mu\omega_{2}\\
-\frac{\mu\omega_{2}}{\sqrt{1-2a}} & 0 & \frac{\mu\omega_{2}}{\sqrt{1+2a}} &
-{\sqrt{2}}\mu\omega_{2}%
\end{array}
\right)  \left(
\begin{array}
[c]{c}%
u_{1}\\
u_{2}\\
u_{3}\\
u_{4}%
\end{array}
\right)
\eeq
where $\omega_{1}$, $\omega_{2}$ and $a$ are given by
\beq
\omega_{1}=\frac{1}{\sqrt{\lam_{1}^{2}\cos^{2} \frac{\theta}%
{2}  +\lam_{2}^{2}\sin^{2} \frac{\theta}{2}  }%
}, \quad \omega_{2}=\frac{1}{\sqrt{\lam_{1}^{2}\sin^{2}  \frac{\theta}%
{2}  +\lam_{2}^{2}\cos^{2}  \frac{\theta}{2}  }}, \quad
a=-\frac{1}{4}\left(  \lam_{1}^{2}-\lam_{2}^{2}\right)  \sin
\theta\omega_{1}\omega_{2}.
\eeq
By showing that 
the determinant of such transformation is
$\frac{\omega_1 \omega_2}{\lam_{1}\lam_{2}}$, it is
easy to map $P_{\alpha\beta}$ in terms of the new variables
as
\beqar
P_{\alpha\beta}  & =&\frac{1}{\pi^{2}}\lam_{1}\lam_{2}\omega_1 \omega_2
e^{-2\left\vert \alpha\right\vert ^{2}-2\left\vert \beta\right\vert
^{2}-\alpha^{2}-\alpha^{\ast2}-\beta^{2}-\beta^{\ast2}}\ \
 \int_{-\infty}^{+\infty}  du_{2}du_{4}\ \  e^{-u_{2}^{2} -u_{4}^{2}} \nonumber\\
&& \times \int_{-\infty}^{+\infty} du_{1} \ e^{-u_{1}^{2}+\frac{\sqrt{2}}%
{\sqrt{1-2a}}\left[ \lam_{1}\left(  \omega_{1}\cos  \frac{\theta
}{2}  +\omega_{2}\sin  \frac{\theta}{2}  \right)  \left(
\alpha+\alpha^{\ast}\right)  +\lam_{2}\left(  \omega_{1}\sin
\frac{\theta}{2}  -\omega_{2}\cos \frac{\theta}{2}
\right)  \left(  \beta+\beta^{\ast}\right)  \right]  u_{1}}  \\
&& \times  \int_{-\infty}^{+\infty} du_{3}\   e^{-u_{3}^{2}+\frac{\sqrt{2}}%
{\sqrt{1+2a}}\left[  \lam_{1}\left(  \omega_{1}\cos  \frac{\theta
}{2}  -\omega_{2}\sin  \frac{\theta}{2}  \right)  \left(
\alpha+\alpha^{\ast}\right)  +\lam_{2}\left(  \omega_{1}\sin
\frac{\theta}{2}  +\omega_{2}\cos  \frac{\theta}{2}
\right)  \left(  \beta+\beta^{\ast}\right)  \right]  u_{3}} \nonumber.
\eeqar
Performing the integration to end up with the result
\beq
P_{\alpha\beta}\left(  \eta,\theta\right)   
 =\frac{1}{\sqrt{2\cosh  2\eta  \sin^{2} \frac{\theta}%
{2} \cos^{2}  \frac{\theta}{2}  +\cos^{4}
\frac{\theta}{2}  +\sin^{4}  \frac{\theta}{2}  }}.
\eeq
This is among the interesting results derived so far in the present work. Indeed,
it shows clearly that 
the purity depends on  the physical parameters $(\eta,\te)$ rather
than the complex displacements $(\al,\be)$ and hereafter it will be
denoted by $P(\eta,\te)$.
Furthermore, the obtained purity is  two parameters
dependent, {which means that it can be controlled easily.
If one requires}
the decoupling case ($\eta=0$), $P({\eta,\te})$ reduces to one
as  expected and therefore there is no entanglement.

To understand better the above results,
we recall that the purity is related to linear entropy by
the simple form
\beq
L= 1-P
\eeq
where $P$ lies in the interval $[0,1]$. Now let us proceed
to plot the purity for a range of
$\eta$ and by considering $\te\in[0,\pi]$. From Figure 1, 
it is clear that the purity, as function of $\eta$, is symmetric with respect to  the {decoupling
case} $\eta=0$. It is maximal
for  $\eta=0$, which really shows that the system is disentangled.
After that it decreases rapidly to reach zero and indicates that the entanglement is
maximal. More importantly,
the purity  becomes constant whenever  $\te$ {takes the value} zero or $\pi$. This behavior of the purity
traced in below tell us that
one can easily play with two parameters to control the degree of entanglement in the present system.

\begin{center}
  \includegraphics[width=4in]{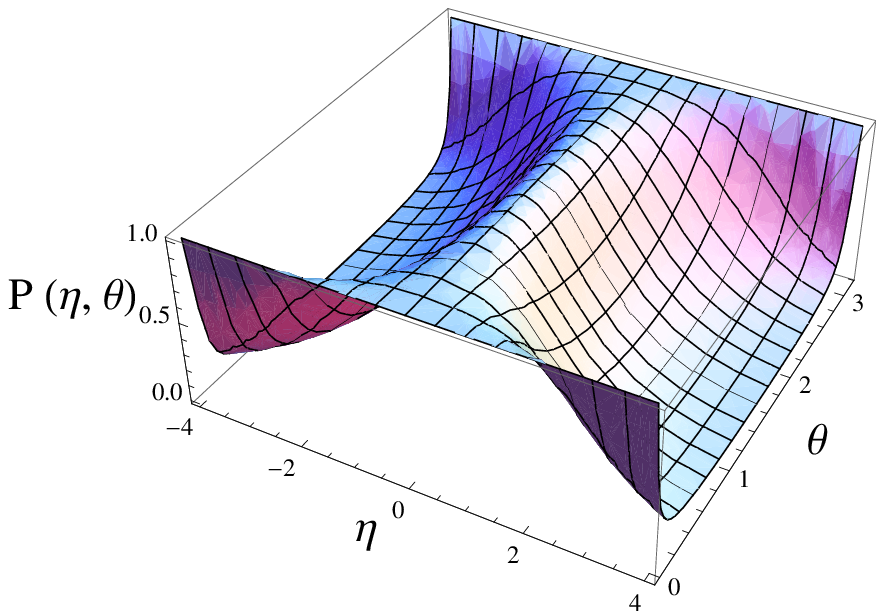}\\
{\sf Figure 1:  {Purity in terms of the coupling parameter $\eta$ and the mixing angle $\te$.}}
\end{center}

 Specifically at $\te=\frac{\pi}{2}$,
we obtain a simple form
\beq\lb{pred}
P\left(  \eta,\theta=\frac{\pi}{2}\right)   
 =\frac{1}{\cosh  \eta}
\eeq
which is one parameter dependent and can be adjusted only
by varying the coupling $\eta$ to control the degree
of the entanglement. To be much more accurate,
we underline such behavior by plotting
(\ref{pred})  in Figure 2:
\begin{center}
  \includegraphics[width=3in]{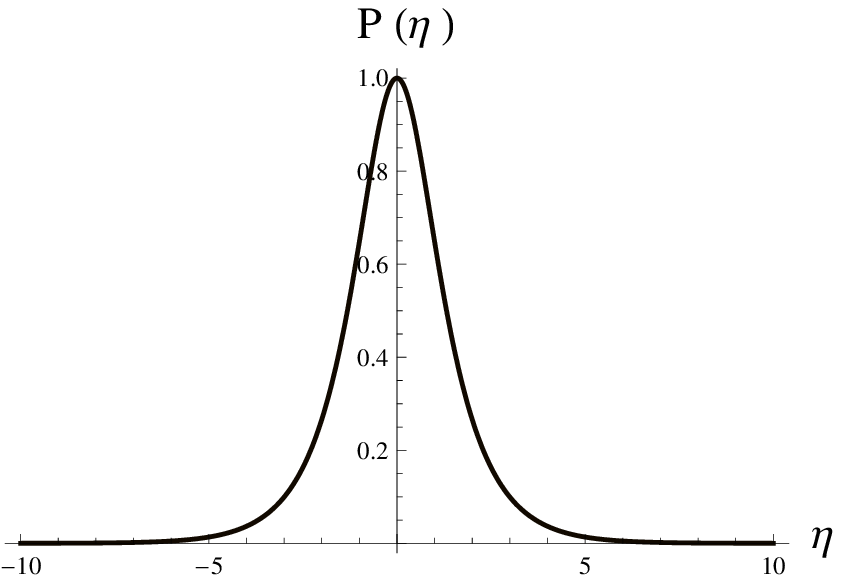}\\
{\sf Figure 2:  {Purity in terms of the coupling parameter $\eta$ for the mixing angle $\te=\frac{\pi}{2}$.}}
\end{center}
From the above figure, one can deduce two interesting conclusions. The first one
tells as that $P(\eta,\te)$ is bounded, i.e. $0 \leq P \leq 1$, as expected. The second one
shows clearly that the purity goes to zero for a strong coupling,
which indicates the entanglement is maximal.

\section{Entanglement in number states} 

To gain more information about the behavior of the present system,
we evaluate the degree of the entanglement between inter states. For this,
we consider the relation inverse to express the number states in terms of
the coherent states. This is
\beq\lb{nscs}
\mid n_1,n_2\rangle=\frac{1}{\sqrt{n_1!n_2!}} 
\left.  \frac
{\partial^{n_1}}{\partial\alpha^{n_1}}\frac{\partial^{n_2}}{\partial\beta^{n_2}%
}e^{\left\vert \alpha\right\vert ^{2}}e^{\left\vert \beta\right\vert ^{2}}%
\mid\alpha,\beta\rangle\right\vert _{\alpha=0,\beta=0}.%
\eeq
In the $y$-representation, (\ref{nscs}) leads to the wavefunction 
\beqar\lb{wf36}
\tilde{\Phi}_{n_1n_2}\left(  y_{1},y_{2}\right)   & =&\tilde\Phi_{n_1n_2}\left(
\mu\cos  \frac{\theta}{2}  X_{1}-\frac{1}{\mu}\sin
\frac{\theta}{2}  X_{2},\mu\sin \frac{\theta}{2}
X_{1}+\frac{1}{\mu}\cos  \frac{\theta}{2}  X_{2}\right) \equiv \tilde\Phi_{n_1n_2}\left(X_1,X_2\right) \nonumber\\
& =&\frac{1}{\sqrt{n_1! n_2!}} 
\left.  \frac{\partial^{n_1}}%
{\partial\alpha^{n_1}}\frac{\partial^{n_2}}{\partial\beta^{n_2}}e^{\frac{\left\vert
\alpha\right\vert ^{2}}{2}}e^{\frac{\left\vert \beta\right\vert ^{2}}{2}}%
\Phi_{\alpha\beta}\left( X_1,X_2\right) 
\right\vert _{\alpha=0,\beta=0}
\eeqar
where $\Phi_{\alpha\beta}\left( X_1,X_2\right)$ is given in (\ref{wf11}). This will be implemented to
study the purity in terms of the number states  and discuss different issues.

Returning back to the purity definition, we have 
\beq
P_{n_1n_2}= \int dX_{1}dX_{1}^{\prime}dX_{2}dX_{2}^{\prime}\tilde{\Phi}%
_{n_1n_2}\left(  X_{1},X_{2}\right)  \tilde{\Phi}_{n_1n_2}^{\ast}\left(
X_{1}^{\prime},X_{2}\right)  \tilde{\Phi}_{n_1n_2}\left(  X_{1}^{\prime}%
,X_{2}^{\prime}\right)  \tilde{\Phi}_{n_1n_2}^{\ast}\left(  X_{1},X_{2}^{\prime
}\right).
\eeq
Using (\ref{wf36}) to obtain the form
\beqar
P_{n_1n_2}
& =&\int dX_{1}dX_{1}^{\prime}dX_{2}dX_{2}^{\prime}\left(  \frac{1}{{n_1!n_2!}%
}\right)  ^{2} \nonumber\\
&& \times\left.  \tfrac{\partial^{n_1}}{\partial\alpha_{1}^{n_1}}\tfrac
{\partial^{n_2}}{\partial\beta_{1}^{n_2}}e^{\frac{\left\vert \alpha_{1}\right\vert
^{2}}{2}}e^{\frac{\left\vert \beta_{1}\right\vert ^{2}}{2}}\Phi_{\alpha
_{1}\beta_{1}}\left(  \mu\cos \frac{\theta}{2} X_{1}-\frac
{1}{\mu}\sin \frac{\theta}{2}  X_{2},\mu\sin  \frac
{\theta}{2}  X_{1}+\frac{1}{\mu}\cos  \frac{\theta}{2}
X_{2}\right)  \right\vert _{\alpha_{1},\beta_{1}=0}\nonumber\\
&& \times\left.  \tfrac{\partial^{n_1}}{\partial\alpha_{2}^{\ast n_1}}%
\tfrac{\partial^{n_2}}{\partial\beta_{2}^{\ast n_2}}e^{\frac{\left\vert \alpha
_{2}\right\vert ^{2}}{2}}e^{\frac{\left\vert \beta_{2}\right\vert ^{2}}{2}%
}\Phi_{\alpha_{2} \beta_{2}}^{\ast}\left(  \mu\cos  \frac{\theta}%
{2}  X_{1}^{\prime}-\frac{1}{\mu}\sin \frac{\theta}{2}
X_{2},\mu\sin  \frac{\theta}{2}  X_{1}^{\prime}+\frac{1}{\mu}%
\cos \frac{\theta}{2}  X_{2}\right)  \right\vert _{\alpha
_{2}^{\ast},\beta_{2}^{\ast}=0}\\
&& \times\left.  \tfrac{\partial^{n_1}}{\partial\alpha_{3}^{n_1}}\tfrac
{\partial^{n_2}}{\partial\beta_{3}^{n_2}}e^{\frac{\left\vert \alpha_{3}\right\vert
^{2}}{2}}e^{\frac{\left\vert \beta_{3}\right\vert ^{2}}{2}}\Phi_{\alpha
_{3} \beta_{3}}\left(  \mu\cos \frac{\theta}{2}  X_{1}^{\prime
}-\frac{1}{\mu}\sin  \frac{\theta}{2}  x_{2}^{\prime},\mu
\sin  \frac{\theta}{2}  X_{1}^{\prime}+\frac{1}{\mu}\cos
\frac{\theta}{2}  x_{2}^{\prime}\right)  \right\vert _{\alpha_{3}%
,\beta_{3}=0}\nonumber\\
& &\times\left.  \tfrac{\partial^{n_1}}{\partial\alpha_{4}^{\ast _1}}%
\tfrac{\partial^{n_2}}{\partial\beta_{4}^{\ast n_2}}e^{\frac{\left\vert \alpha
_{4}\right\vert ^{2}}{2}}e^{\frac{\left\vert \beta_{4}\right\vert ^{2}}{2}%
}\Phi_{\alpha_{4} \beta_{4}}^{\ast}\left(  \mu\cos  \frac{\theta}%
{2}  x_{1}-\frac{1}{\mu}\sin  \frac{\theta}{2}
x_{2}^{\prime},\mu\sin \frac{\theta}{2}  X_{1}+\frac{1}{\mu}%
\cos \frac{\theta}{2}  X_{2}^{\prime}\right)  \right\vert
_{\alpha_{4}^{\ast},\beta_{4}^{\ast}=0}.\nonumber
\eeqar
After some algebra, we show that the purity takes the form
\beqar
P_{n_1n_2}   &=& 
\left(  \frac
{\lam_1 \lam_2}{\pi n_1!n_2!}\right)  ^{2}%
{\textstyle\prod\limits_{i=1}^{4}}
\frac{\partial^{n_1}}{\partial\alpha_{i}^{n_1}}\frac{\partial^{n_2}}{\partial
\beta_{i}^{n_2}}\int dX_{1}dX_{1}^{\prime}dX_{2}dX_{2}^{\prime} \ \
 e^{-\frac{1}{2}\left(\alpha_{1}^{2} + \beta_{1}^{2} +
\alpha_{2}^{2}+
\beta_{2}^{2} + \alpha_{3}^{2} + \beta_{3}^{2} +
\alpha_{4}^{2} + \beta_{4}^{2}\right)} \nonumber\\
 &&  e^{-\frac{\lam_{1}^{2}}{2}\left[\left(  \mu\cos \frac{\theta
}{2}  X_{1}-\frac{1}{\mu}\sin \frac{\theta}{2}
X_{2}\right)  ^{2}
+\left(  \mu\cos \frac{\theta}{2}
X_{1}^{\prime}-\frac{1}{\mu}\sin \frac{\theta}{2}  X_{2}%
^{\prime}\right)  ^{2} + \left(  \mu\cos
\frac{\theta}{2}  X_{1}^{\prime}-\frac{1}{\mu}\sin \frac{\theta
}{2} X_{2}\right)  ^{2}
+\left(  \mu\cos
\frac{\theta}{2}  X_{1}-\frac{1}{\mu}\sin \frac{\theta}%
{2}  X_{2}^{\prime}\right)  ^{2}\right]} \nonumber\\
&& e^{-\frac{\lam_{2}^{2}}{2}\left[\left(  \mu\sin
\frac{\theta}{2}  X_{1}+\frac{1}{\mu}\cos  \frac{\theta}%
{2}  X_{2}\right)  ^{2}
+\left(  \mu\sin \frac{\theta}{2}
X_{1}^{\prime}+\frac{1}{\mu}\cos \frac{\theta}{2}  X_{2}%
^{\prime}\right)  ^{2}
+\left(  \mu\sin \frac{\theta}{2}
X_{1}+\frac{1}{\mu}\cos  \frac{\theta}{2}  X_{2}^{\prime}\right)
^{2}
+\left(  \mu\sin  \frac{\theta}{2}
X_{1}^{\prime}+\frac{1}{\mu}\cos \frac{\theta}{2}  X_{2}
\right)^{2}\right]}\nonumber\\
&& e^{\sqrt{2}\mu\left[\left(  \lam_{1}\left(  \alpha_{1}+\alpha_{4}\right)
\cos  \frac{\theta}{2} +\lam_{2}\left(  \beta_{1}+\beta
_{4}\right)  \sin \frac{\theta}{2} \right)  X_{1}
+\left(  \lam_{1}\left(  \alpha_{2}+\alpha_{3}\right)
\cos \frac{\theta}{2}  +\lam_{2}\left(  \beta_{2}+\beta
_{3}\right)  \sin \frac{\theta}{2}  \right)  X_{1}^{\prime}\right]}\nonumber\\
&& e^{-\frac{\sqrt{2}}{\mu}\left[\left(  \lam_{1}\left(  \alpha_{1}+\alpha
_{2}\right)  \sin  \frac{\theta}{2}  -\lam_{2}\left(
\beta_{2}+\beta_{1}\right)  \cos  \frac{\theta}{2}  \right)
X_{2}
+\left(  \lam_{1}\left(  \alpha_{3}%
+\alpha_{4}\right)  \sin  \frac{\theta}{2}  -\lam_{2}\left(
\beta_{3}+\beta_{4}\right)  \cos  \frac{\theta}{2}  \right)
X_{2}^{\prime}\right]}.
\eeqar
This can be written, in a compact form, as 
\beq\lb{puco}
P_{n_1n_2}= 
\left(  \frac{\lam_1\lam_2}%
{\pi n_1!n_2!}\right)  ^{2}%
{\textstyle\prod\limits_{i=1}^{4}}
\frac{\partial^{n_1}}{\partial\alpha_{i}^{n_1}}\frac{\partial^{n_2}}{\partial
\beta_{i}^{n_2}}\int d^{4}Ze^{-Z^{t}\cdot A\cdot Z+B^{t}\cdot Z+C}%
\eeq
where
$z^{T}=\left(
\begin{array}
[c]{cccc}%
X_{1} & X_{1}^{\prime} & X_{2} & X_{2}^{\prime}%
\end{array}
\right)$,
 the matrix $A$ is given by
\beq
A=\left(
\begin{array}
[c]{cccc}%
A_{11} & 0 & A_{13} & A_{13}\\
0 & A_{11} & A_{13} & A_{13}\\
A_{13} & A_{13} & A_{33} & 0\\
A_{13} & A_{13} & 0 & A_{33}%
\end{array}
\right)
\eeq
such that their components read as
\beq
A_{11}=\mu^2\left(\lam_{1}^{2}\cos^2\frac{\theta}{2}+\lam_{2}^{2}\sin^2 \frac{\theta}{2}\right),\quad
A_{33}=\frac{1}{\mu} \left(\lam_{1}^{2}\sin^2 \frac{\theta}{2} + \lam_{2}^{2} \cos^2 \frac{\theta}{2}\right),\quad
A_{13}=\frac{\lam_{2}^{2}-\lam_{1}^{2}}{4}\sin \theta
\eeq
and the matrix $B$ takes the form
\beq
B=\sqrt{2}\left(
\begin{array}
[c]{c}%
\mu\lam_{1}\left(  \alpha_{1}+\alpha_{4}\right)  \cos \frac{\theta
}{2}  +\mu\lam_{2}\left(  \beta_{1}+\beta_{4}\right)  \sin
\frac{\theta}{2}  \\
\mu\lam_{1}\left(  \alpha_{2}+\alpha_{3}\right)  \cos  \frac{\theta
}{2} +\mu\lam_{2}\left(  \beta_{2}+\beta_{3}\right)  \sin
\frac{\theta}{2}  \\
\frac{1}{\mu}\lam_{2}\left(  \beta_{1}+\beta_{2}\right)  \cos
\frac{\theta}{2}  -\frac{1}{\mu}\lam_{1}\left(  \alpha_{1}%
+\alpha_{2}\right)  \sin  \frac{\theta}{2}  \\
\frac{1}{\mu}\lam_{2}\left(  \beta_{3}+\beta_{4}\right)  \cos
\frac{\theta}{2}  -\frac{1}{\mu}\lam_{1}\left(  \alpha_{3}%
+\alpha_{4}\right)  \sin  \frac{\theta}{2}
\end{array}
\right).
\eeq

To go further in evaluating the purity, we perform a method to simplify our calculation. This can be done
by introducing the
change of variables
\beq
\left(
\begin{array}
[c]{c}%
X_{1}\\
X_{1}^{\prime}\\
X_{2}\\
X_{2}^{\prime}%
\end{array}
\right)  =\left(
\begin{array}
[c]{cccc}%
\frac{\omega_{1}}{2\mu\sqrt{1-2a}} & \frac{\sqrt{2}}{2\mu}\omega_{1} &
\frac{\omega_{1}}{2\mu\sqrt{1+2a}} & 0\\
\frac{\omega_{1}}{2\mu\sqrt{1-2a}} & -\frac{\sqrt{2}}{2\mu}\omega_{1} &
\frac{\omega_{1}}{2\mu\sqrt{1+2a}} & 0\\
-\frac{\mu\omega_{2}}{2\sqrt{1-2a}} & 0 & \frac{\mu\omega_{2}}{2\sqrt{1+2a}} &
\frac{\sqrt{2}}{2}\mu\omega_{2}\\
-\frac{\mu\omega_{2}}{2\sqrt{1-2a}} & 0 & \frac{\mu\omega_{2}}{2\sqrt{1+2a}} &
-\frac{\sqrt{2}}{2}\mu\omega_{2}%
\end{array}
\right)  \left(
\begin{array}
[c]{c}%
x_{1}\\
x_{2}\\
x_{3}\\
x_{4}%
\end{array}
\right)
\eeq
where the corresponding measure is 
$dX_{1}dX_{1}^{\prime}dX_{2}dX_{2}^{\prime}=Jdx_{1}dx_{1}dx_{2}dx_{2}$
and 
the Jacobian reads as 
\beq
J=\frac{1}{\lam_{1}\lam_{2}\sqrt{\left(  \lam_{1}^{2}\cos
^{2} \frac{\theta}{2}  +\lam_{2}^{2}\sin^{2}
\frac{\theta}{2}  \right)  \left(  \lam_{1}^{2}\sin^{2}
\frac{\theta}{2}  +\lam_{2}^{2}\cos^{2} \frac{\theta}%
{2}  \right)  }}.%
\eeq
This performance allows us to map (\ref{puco}) as
\beq\lb{newp}
P_{n_1n_2}= 
\left(  \frac
{\lam_1 \lam_2}{\pi n_1!n_2!}\right)  ^{2} J%
{\textstyle\prod\limits_{i=1}^{4}}
\frac{\partial^{n_1}}{\partial\alpha_{i}^{n_1}}\frac{\partial^{n_2}}{\partial
\beta_{i}^{n_2}}e^{-\frac{1}{2}\left(  \alpha_{i}^{2}+\beta_{i}^{2}\right)
}\int d^{4}Qe^{-Q^{2}+D^{t}Q}%
\eeq
where $Q^{t}=\left(
\begin{array}
[c]{cccc}%
x_{1} & x_{2} & x_{3} & x_{4}%
\end{array}
\right)  $ and $D^{t}$ is the transpose of $D$, such as
\beq
D=\sqrt{2}\left(
\begin{array}
[c]{c}%
\frac{\omega_{1}\cos     \frac{\theta}{2}    +\omega_{2}\sin
\frac{\theta}{2}  }{2\sqrt{1-2a}}\lam_{1}\left(  \alpha_{1}%
+\alpha_{2}+\alpha_{3}+\alpha_{4}\right)  +\frac{\omega_{1}\sin
\frac{\theta}{2} -\omega_{2}\cos \frac{\theta}{2}
}{2\sqrt{1-2a}}\lam_{2}\left(  \beta_{1}+\beta_{2}+\beta_{3}+\beta
_{4}\right)  \\
\frac{\sqrt{2}}{2}\omega_{1}\lam_{1}\left(  \alpha_{1}+\alpha_{4}%
-\alpha_{2}-\alpha_{3}\right)  \cos  \frac{\theta}{2}
+\frac{\sqrt{2}}{2}\omega_{1}\lam_{2}\left(  \beta_{1}+\beta_{4}-\beta
_{2}-\beta_{3}\right)  \sin  \frac{\theta}{2}  \\
\frac{\omega_{1}\cos  \frac{\theta}{2}  -\omega_{2}\sin\left(
\frac{\theta}{2}\right)  }{2\sqrt{1+2a}}\lam_{1}\left(  \alpha_{1}%
+\alpha_{4}+\alpha_{2}+\alpha_{3}\right)  +\frac{\omega_{1}\sin
\frac{\theta}{2}  +\omega_{2}\cos \frac{\theta}{2}
}{2\sqrt{1+2a}}\lam_{2}\left(  \beta_{1}+\beta_{4}+\beta_{2}+\beta
_{3}\right)  \\
-\frac{\sqrt{2}}{2}\omega_{2}\lam_{1}\left(  \alpha_{1}+\alpha_{2}%
-\alpha_{3}-\alpha_{4}\right)  \sin  \frac{\theta}{2}
+\frac{\sqrt{2}}{2}\omega_{2}\lam_{2}\left(  \beta_{1}+\beta_{2}-\beta
_{3}-\beta_{4}\right)  \cos  \frac{\theta}{2}
\end{array}
\right).
\eeq
Since the above integral is Gaussian, then after some algebra
we end with the form 
\beqar\lb{anpu}
 P_{n_1n_2} &=& 
 \left(  \frac
{\lam_1 \lam_2}{\pi n_1!n_2!}\right)  ^{2}J
{\textstyle\prod\limits_{i=1}^{4}}
\frac{\partial^{n_1}}{\partial\alpha_{i}^{n_1}}\frac{\partial^{n_2}}{\partial
\beta_{i}^{n_2}}\exp\left[  \frac{u}{\rho}\alpha_{1}^{2}+\frac{2v}{\rho}%
\alpha_{1}\alpha_{2}-\frac{2u}{\rho}\alpha_{1}\alpha_{3}+\frac{2w}{\rho}%
\alpha_{1}\alpha_{4}+\frac{2s}{\rho}\alpha_{1}\beta_{1}\right.  \nonumber\\
&& -\frac{2t}{\rho}\alpha_{1}\beta_{2}-\frac{2s}{\rho}\alpha_{1}\beta_{3}%
+\frac{2t}{\rho}\alpha_{1}\beta_{4}+\frac{u}{\rho}\alpha_{2}^{2}+\frac
{2w}{\rho}\alpha_{2}\alpha_{3}-\frac{2u}{\rho}\alpha_{2}\alpha_{4}-\frac
{2t}{\rho}\alpha_{2}\beta_{1}+\frac{2s}{\rho}\alpha_{2}\beta_{2}\nonumber\\
&& +\frac{2t}{\rho}\alpha_{2}\beta_{3}-\frac{2s}{\rho}\alpha_{2}\beta_{4}%
+\frac{u}{\rho}\alpha_{3}^{2}+\frac{2v}{\rho}\alpha_{3}\alpha_{4}-\frac
{2s}{\rho}\alpha_{3}\beta_{1}+\frac{2t}{\rho}\alpha_{3}\beta_{2}+\frac
{2s}{\rho}\alpha_{3}\beta_{3}-\frac{2t}{\rho}\alpha_{3}\beta_{4}\\
&& +\frac{u}{\rho}\alpha_{4}^{2}+\frac{2t}{\rho}\alpha_{4}\beta_{1}-\frac
{2s}{\rho}\alpha_{4}\beta_{2}-\frac{2t}{\rho}\alpha_{4}\beta_{3}+\frac
{2s}{\rho}\alpha_{4}\beta_{4}-\frac{u}{\rho}\beta_{1}^{2}+\frac{2w}{\rho}%
\beta_{1}\beta_{2}\nonumber\\
&& \left.  +\frac{2u}{\rho}\beta_{1}\beta_{3}+\frac{2v}{\rho}\beta_{1}\beta
_{4}-\frac{u}{\rho}\beta_{2}^{2}+\frac{2v}{\rho}\beta_{2}\beta_{3}+\frac
{2u}{\rho}\beta_{2}\beta_{4}-\frac{u}{\rho}\beta_{3}^{2}+\frac{2w}{\rho}%
\beta_{3}\beta_{4}-\frac{u}{\rho}\beta_{4}^{2}\right]\nonumber
\eeqar
where we have set the involved parameters as
\beqar
\rho &=& 4\frac{mk}{\hbar^{2}}\left[  2\cosh   (2\eta)   +\cot^{2}
\frac{\theta}{2}   +\tan^{2}  \frac{\theta}{2}   \right], \qquad
u=\frac{2mk}{\hbar^{2}}\sinh  2\eta  \nonumber\\
v&=&\frac{2mk}{\hbar^{2}}\left[  \cosh  (2\eta)   +\tan^{2}
\frac{\theta}{2}   \right], \qquad
w=\frac{2mk}{\hbar^{2}}\left[  \cosh   (2\eta)   +\cot^{2}
\frac{\theta}{2}   \right]\\
t&=&4\frac{mk}{\hbar^{2}}\frac{\cosh  \eta   }{\sin
\theta   }, \qquad
s=-4\frac{mk}{\hbar^{2}}\sinh   \eta   \frac{\cos
\theta   }{\sin   \theta  \nonumber }.%
\eeqar

We are still looking for the final form of the purity, which can be obtained by calculating the partial derivatives.
These can be performed in different ways and may be it is easier to proceed step by step. Indeed, we factorize
the exponential function and then map each factor into a series expansion. This operation has been postponed to
Appendix A and the yielded result is
\beq\lb{fpur}
P_{n_1n_2}\left(  \eta,\theta\right)  =\tfrac{2\left(  \frac{2}{\rho}\right)
^{2(n_1+n_2)} (n_1!n_2!)^2}{\sin\left(  \theta\right)  \sqrt{2\cosh2\eta+\tan^{2}\left(
\frac{\theta}{2}\right)  +\cot^{2}\left(  \frac{\theta}{2}\right)  }}%
{\textstyle\sum\limits_{i+j+k+l+r=2(n_1+n_2)}}
C_{n_1n_2}\left(  i,j,k,l,r\right)  u^{i}v^{j}w^{k}t^{l}s^{r}%
\eeq
where the coefficients $C_{n_1n_2}$ are given by
\beqar
C_{n_1n_2}   & =&\left(
{\textstyle\prod\limits_{e=1}^{4}}
{\textstyle\sum\limits_{i_{e}=0}^{i_{e-1}}}
\right)  \left(
{\textstyle\prod\limits_{e=1}^{3}}
{\textstyle\sum\limits_{j_{e}=0}^{j_{e-1}}}
\right)  \left(
{\textstyle\prod\limits_{e=1}^{3}}
{\textstyle\sum\limits_{k_{e}=0}^{k_{e-1}}}
\right)  \left(
{\textstyle\prod\limits_{e=1}^{7}}
{\textstyle\sum\limits_{l_{e}=0}^{l_{e-1}}}
\frac{1}{\left(  l_{e-1}-l_{e}\right)  !}\right)
\left(
{\textstyle\prod\limits_{e=1}^{7}}
{\textstyle\sum\limits_{r_{e}=0}^{r_{e-1}}}
\frac{1}{\left(  r_{e-1}-r_{e}\right)  !}\right) \nonumber\\
&&
{\frac{ 2^{-i_{4}}
(-1)^{l_{1}-l_{3}+l_{4}-l_{5}+l_{6}-l_{7} + r-r_{1}+r_{3}-r_{5}+r_{6}-r_{7}+ i_2-c_1-c_2}}
{{ \left(  i-i_{1}\right)!\left(  i_{1}-i_{2}\right)  !\left(  i_{2}-i_{3}\right)  !
\left(  i_{3}-i_{4}\right)
! l_{7}! r_{7}! c_3!c_4! c_5! c_6!c_7! c_8!c_9! c_{10}! }}}.
\eeqar
It is clear that the final {form of the purity is actually
 only depending on two parameters, i.e.  $\eta$ and $\te$.} 
 On the other hand, {it is easy to check that} $P_{n_1n_2}$
is symmetric under the change of the quantum numbers $n_1$ and $n_2$.


\section{Two special cases}%

To be much more accurate let is illustrate some particular cases. With these
we will be able to get more information from the above purity about the degree
of entanglement.
In the beginning, let us choose the configuration $(n_1=0,n_2=1)$, which means
that we are considering now the entanglement between the ground state
of the first oscillator
and the first excited state of the second one. In this case,
(\ref{fpur}) reduces to the form
\beq
P_{01}\left(  \eta,\theta\right)     =\tfrac{2\left(  \frac{2}{\rho}\right)
^{2}}{\sin  \theta  \sqrt{2\cosh(2\eta)+\tan^{2}\frac
{\theta}{2}  +\cot^{2} \frac{\theta}{2}  }}\
{\textstyle\sum\limits_{l+r+j+k+i=2}}
C_{01}\left(  i,j,k,l,r\right)  u^{i}v^{j}w^{k}t^{l}s^{r}
\eeq
which can be evaluated to obtain
\beq
P_{01}\left(  \eta,\theta\right)     =
\tfrac{2\left(  \frac{2}{\rho}\right)  ^{2}}{\sin  \theta
\sqrt{2\cosh(2\eta)+\tan^{2} \frac{\theta}{2}  +\cot^{2}
\frac{\theta}{2}  }}\left(  u^{2}+v^{2}+w^{2}\right)
\eeq
and after replacing different parameters, one gets the final result
\beq\lb{p01}
P_{01}\left(  \eta,\theta\right)     =
\frac{3\cosh\left(  4\eta\right)  +4\left(  \tan^{2} \frac{\theta
}{2}  +\cot^{2}  \frac{\theta}{2}  \right)  \cosh\left(
2\eta\right)  +2\tan^{4} \frac{\theta}{2}  +2\cot^{4}
\frac{\theta}{2}  +1}{\sin \theta  \left(  2\cosh
(2\eta) +\tan^{2} \frac{\theta}{2}  +\cot^{2}
\frac{\theta}{2}  \right)  ^{\frac{5}{2}}}.
\eeq
This is a nice form, which can be worked more since it is only function of two
physical parameters $\eta$ and $\te$. Indeed, we plot it in Figure 3:
\begin{center}
  \includegraphics[width=4in]{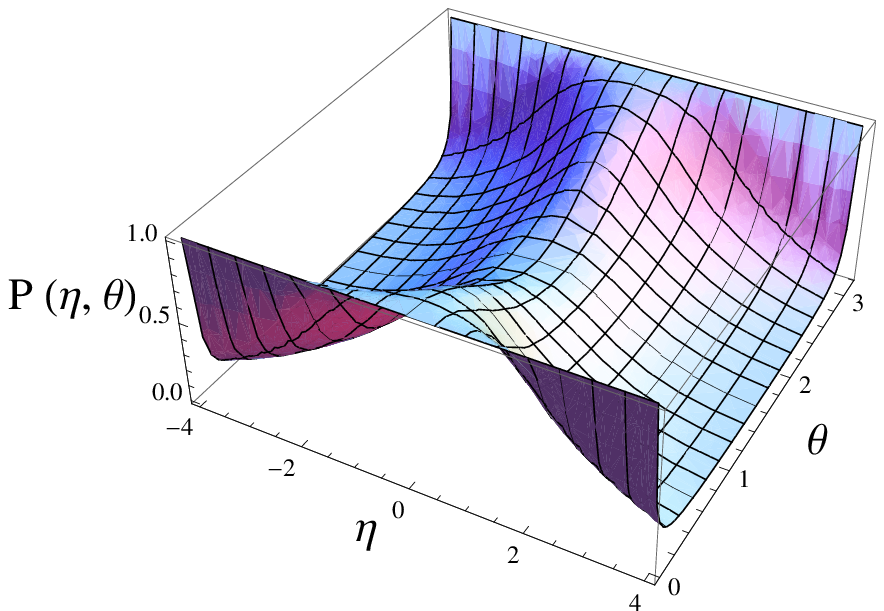}\\
{\sf Figure 3:  {Purity $P_{01}$ as function of the coupling parameter $\eta$
and  mixing angle $\te$ for the quantum numbers $(n_1=0,n_2=1) $.}}
\end{center}
Here we have the same conclusion as in Figure 1 except that the
present plot is showing some deformation at the point
$\eta=0$. Otherwise, for certain values
of $\te$ the purity is not always holding a maximum value
at $\eta=0$. More precisely, at this point it decreases to reach $1/2$ at $\te=\frac{\pi}{2}$ and
then increases to attends 1 at  $\te={\pi}$. This is because in the present case
the masses are equal and the same conclusion is obtained in \cite{harshman}.

Now
let us look at some interesting situations by fixing the mixing angle $\theta$
and varying
the coupling parameter $\eta$.
In particular when $\te=\frac{\pi}{2}$, $P_{01}$ reduces
to the form
\beq
P_{01}\left(  \eta,\theta=\frac{\pi}{2}\right)     =
\frac{3\cosh (4\eta) + 8\cosh(2\eta) +5} {32 \cosh^5{\eta}}.
\eeq
This can be plotted to obtain Figure 4: 
\begin{center}
  \includegraphics[width=4in]{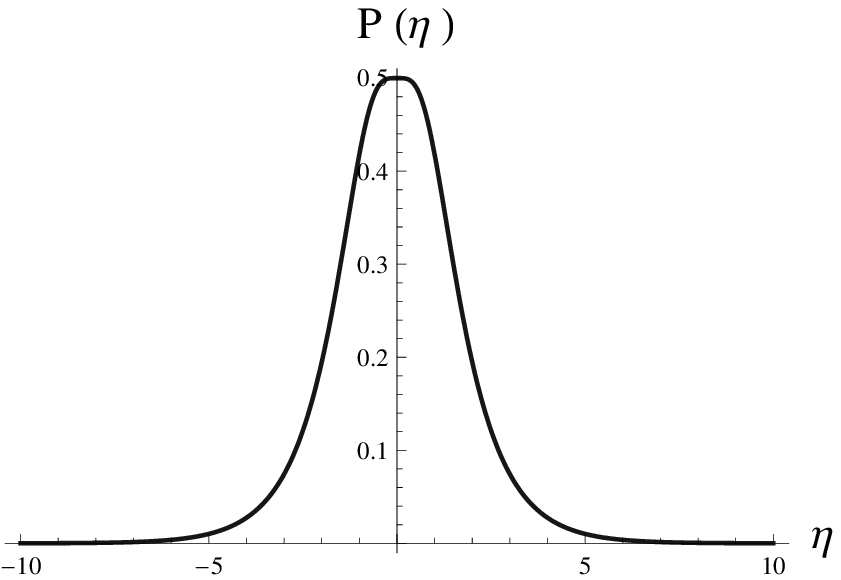}\\
{\sf Figure 4:  {Purity $P_{01}$ as function of $\eta$ measuring
the entanglement between the ground state $n_1=0$ and the first excited state $n_2=1$
for $\te=\frac{\pi}{2}$.}}
\end{center}
Compared to Figure 2, we notice that the behavior of the purity in terms of
the coupling parameter $\eta$ is large. As long as $\eta$ is
large the entanglement is going to hold the
maximum value. It shows clearly the role playing by $\eta$ and thus
 allows an easy control of the degree of the entanglement. This
may give some hint about an experiment realization of the present
case.

Now let us look at the case of the entanglement between
the two first excited states of the two oscillators, i.e. $n_1=n_2=1$. This
result gives
\beq
P_{11}\left(  \eta,\theta\right)  =\tfrac{2\left(  \frac{2}{\rho}\right)
^{4}}{\sin  \theta  \sqrt{2\cosh(2\eta)+\tan^{2}  \frac
{\theta}{2}  +\cot^{2}  \frac{\theta}{2}  }}\
{\textstyle\sum\limits_{i+j+k+l+r=4}}
C_{11}\left(  i,j,k,l,r\right)  u^{i}v^{j}w^{k}t^{l}s^{r}%
\eeq
after lengthy but simple calculations, we find%
\beqar\lb{p11f}
P_{11}  &=&\tfrac{2\left(  \frac{2}{\rho}\right)  ^{4}}{\sin
\theta  \sqrt{2\cosh (2\eta)  +\tan^{2} \frac{\theta
}{2}  +\cot^{2}  \frac{\theta}{2}  }}\\
&&\times \left(  u^{4}%
+v^{4}+w^{4}+2^{2}s^{4}+2^{2}t^{4}+2^{4}s^{2}t^{2}+2u^{2}v^{2}+2u^{2}%
w^{2}+2v^{2}w^{2}  +2^{4}ustv
-2^{4}ustw
\right.  \nonumber\\
 &&\left.-2^{4}u^{2}s^{2}-2^{4}u^{2}t^{2}-2^{4}t^{2}%
w^{2}-2^{4}s^{2}w^{2}-2^{4}t^{2}v^{2}-2^{4}s^{2}v^{2}+2^{3}vws^{2}%
+2^{3}vwt^{2}\right).\nonumber
\eeqar
Finally, we obtain
\beqar\lb{p11}
P_{11}\left(  \eta,\theta\right)    & =& \frac{1}{4\sin \theta
\left[  2\cosh  (2\eta)  +\tan^{2}  \frac{\theta}{2}
+\cot^{2}  \frac{\theta}{2}  \right]  ^{\frac{9}{2}}} 
 \left[
9\cosh\left(  8\eta\right)  +16\left(  \tan^{2}  \frac{\theta}%
{2}  +\cot^{2}  \frac{\theta}{2}  \right)  \cosh
6\eta  \right.  \nonumber\\
&& +\left(  96\tan^{4}  \frac{\theta}{2}  +96\cot^{4}
\frac{\theta}{2}  -36\right)  \cosh\left(  4\eta\right)  +240\left(
\tan^{2}  \frac{\theta}{2}  +\cot^{2}  \frac{\theta}%
{2}  \right)  \cosh\left(  2\eta\right)  \nonumber\\
&& \left.  +8\tan^{8}  \frac{\theta}{2}  +8\cot^{8}
\frac{\theta}{2} -64\tan^{4}  \frac{\theta}{2}
-64\cot^{4}  \frac{\theta}{2}  +459\right].
\eeqar
Comparing this with (\ref{p01}), we notice that the numerator of both
of them is containing a hyperbolic cosine function of a even number
of coupling parameter $\eta$ and the denominators
are power of $\cosh{\eta}$. To go further, we plot (\ref{p11f}) in
Figure 5:
\begin{center}
  \includegraphics[width=4in]{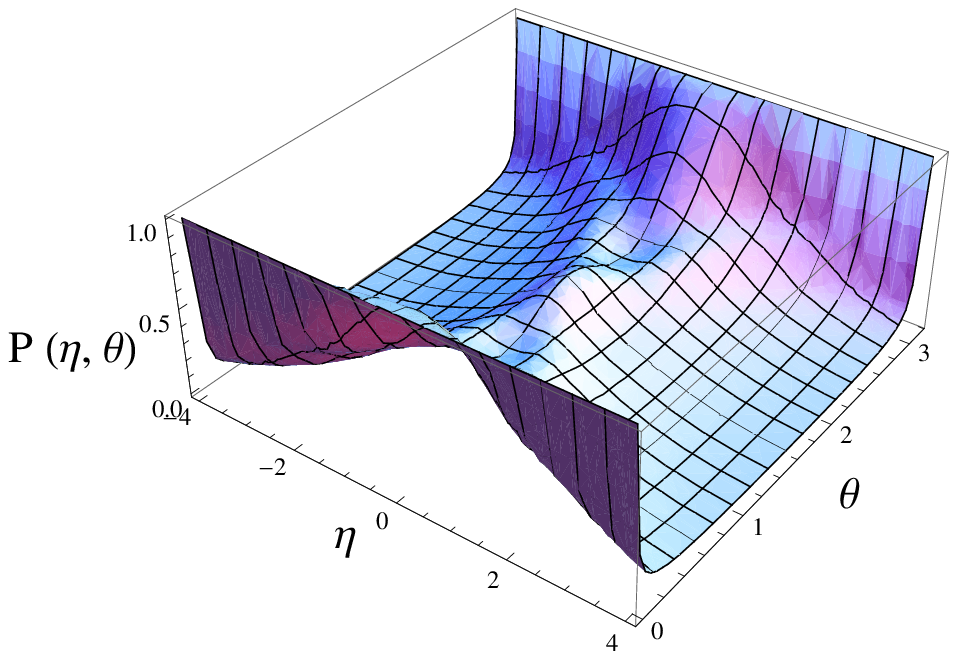}\\
{\sf Figure 5:  {Purity $P_{11}$ as function of the coupling parameter $\eta$
and  mixing angle $\te$ for the quantum numbers $(n_1=1,n_2=1) $.}}
\end{center}
Clearly, we see that
for certain values
of $\te$ the purity is not always holding a maximum value
at decoupling case, i.e $\eta=0$. At this point,  the purity decreases to reach $1/2$ at $\te=\frac{\pi}{2}$ and
then increases to attends 1 at  $\te={\pi}$.

Furthermore,  (\ref{p11})
can be worked much more to underline its behavior. The simplest way to do
so is to fix the mixing angle $\te$ and play with  the coupling parameter
$\eta$.
For instance, by requiring  $\te=\frac{\pi}{2}$ we end up with the form
\beq\lb{p11t}
P_{11}\left(  \eta,\theta\right)    =
\frac{9\cosh(8\eta) + 32\cosh(6\eta) + 156 \cosh(4\eta) + 480 \cosh(2\eta) + 347}
{2048 \cosh^9{\eta}}.
\eeq
This shows clearly that $P_{11}\left(  \eta,\theta\right)$ is one parameter dependent
and therefore it can be manipulated easily. For more precision,
we plot (\ref{p11t}) in Figure 6:
\begin{center}
  \includegraphics[width=4in]{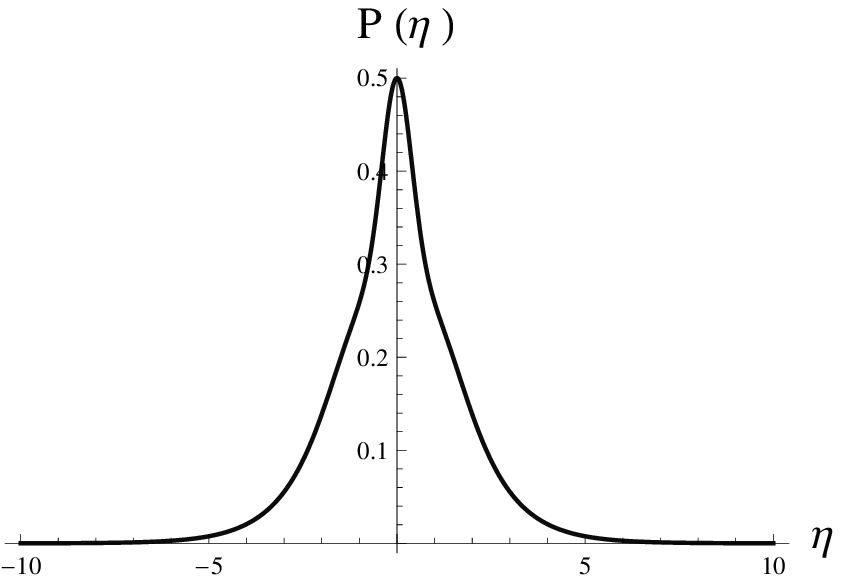}\\
{\sf Figure 6:  {Purity $P_{11}$ as function of $\eta$ measuring
the entanglement between the first exited  states $(n_1=1,n_2=1)$
for $\te=\frac{\pi}{2}$.}}
\end{center}
This is showing a difference with respect to Figure 4. It is clear that
as long as $\eta$ is small the purity increases rapidly to reach its maximal
value. Also it decreases rapidly to attend zero for large $\eta$, which
 means that the system is strongly entangled.

\section{Conclusion}

The present work is devoted to study the entanglement of
two coupled harmonic oscillators by adopting a new approach.
For this, a Hamiltonian describing the system is considered
and an unitary transformation is introduced. With this latter,
the corresponding solutions of the energy spectrum are obtained
in terms of the coupling parameter $\eta$ and the mixing angle
$\te$. It is clearly seen that when $\eta=0$, the system becomes
decoupled and therefore nothing new except harmonic oscillator
in two dimensions.

To study the entanglement of the present system, we have introduced
the purity function to evaluate its degree. In the beginning,
we have realized the corresponding coherent states by using the standard
method based on the displacement operator. These are used to determine explicitly the form of the
purity in terms of the physical parameters $\eta$ and $\te$.
Also, the obtained result
confirmed the range of the purity that is $0\leq P\leq 1$. Moreover,
we have clearly shown that   purity is easy to control and can
 also be cast in a simple form when we fix  $\te=\frac{\pi}{2}$.
 In such case
the purity is obtained  as the inverse of the hyperbolic function
$\cosh\eta$ and the disentanglement simply corresponds to switching off
 $\eta$.

Subsequently, we have used the relation inverse between the number
of states and the coherent states to determine the purity.
After making different changes of variable, we have got a tractable Gaussian
form, which was integrated easily. The final result showed
that the purity is two parameters {dependent.} 
This allowed us
to illustrate our finding by restricting ourselves to
two particular cases. In the first configuration, we have considered
the entanglement between the ground state and excited state, i.e.
$(n_1=0,n_2=1)$ where the purity is exactly obtained.
In the second configuration we studied
the entanglement between  the states $(n_1=1,n_2=1)$. In both cases,
we have analyzed the case where $\te=\frac{\pi}{2}$, which showed that
a strong dependence of the purity to the hyperbolic cosine function
of even coupling parameter.

{On the other hand, the system of two coupled
oscillators can serve as an analog computer for many of the
physical theories and models. Therefore,
one can extend the method developed here to study the entanglement in
other interesting systems those illustrating the Feynman's rest. Furthermore,
one immediate
extension is to
consider} the case of 
a coupled systems submitted to an external magnetic field. This work and related matter
 are actually under consideration.

\section*{Acknowledgments}

The authors acknowledge {the financial} support by  King Faisal University. The present work is done under
 Project Number: 110135, "Quantum information and Entangled Nano Electron Systems".
The authors would like to thank E.B. Choubabi for the  numerical help
{and are indebted to the referee
for his constructive comment.}

\section*{{\bf Appendix A: Final form of purity}}

In this appendix, we show how to derive the final form of the purity given in
(\ref{fpur}). Indeed
from (\ref{anpu}), we obtain the result
\begin{align}
P_{n_1n_2} &  =\left(\frac{\lam_{1}\lam_{2}}{\pi n_1!n_2!}\right)^2 J
{\textstyle\sum\limits_{i,j,k,l,r=0}^{\infty}}
\left(
{\textstyle\prod\limits_{e=1}^{11}}
{\textstyle\sum\limits_{i_{e}=0}^{i_{e-1}}}
\frac{1}{\left(  i_{e-1}-i_{e}\right)  !}\right)  \left(
{\textstyle\prod\limits_{e=1}^{3}}\left[
{\textstyle\sum\limits_{j_{e}=0}^{j_{e-1}}}
\frac{1}{\left(  j_{e-1}-j_{e}\right)  !}
{\textstyle\sum\limits_{k_{e}=0}^{k_{e-1}}}
\frac{1}{\left(  k_{e-1}-k_{e}\right)  !}\right]\right) \nonumber \\
 &\left(
{\textstyle\prod\limits_{e=1}^{7}}\left[
{\textstyle\sum\limits_{l_{e}=0}^{l_{e-1}}}
\frac{1}{\left(  l_{e-1}-l_{e}\right)  !}
{\textstyle\sum\limits_{r_{e}=0}^{r_{e-1}}}
\frac{1}{\left(  r_{e-1}-r_{e}\right)  !}\right]\right)
\left(  \frac{u}{\rho}\right)  ^{i}\left(  \frac{2v}{\rho}\right)
^{j}\left(  \frac{2w}{\rho}\right)  ^{k}\left(  \frac{2t}{\rho}\right)
^{l}\left(  \frac{2s}{\rho}\right)  ^{r}\frac{2^{i-i_{4}}}{i_{11}!j_{3}%
!k_{3}!l_{7}!r_{7}!}\nonumber\\
&\left(  \tfrac{\partial^{n_1}}{\partial\alpha_{1}^{n_1}} \alpha_{1}^{a_1}\right)
\left(  \tfrac{\partial^{n_1}}{\partial\alpha_{2}^{n_1}}\alpha_{2}^{a_2}\right)
\left(  \tfrac{\partial^{n_1}}{\partial\alpha_{3}^{n_1}}\alpha_{3}^{a_3}\right)
\left(  \tfrac{\partial^{n_1}}{\partial\alpha_{4}^{n_1}}\alpha_{4}^{a_4}\right)
 \left(  \tfrac{\partial^{n_2}}{\partial\beta_{1}^{n_2}}\beta_{1}^{a_5}\right)
 \left(  \tfrac{\partial^{n_2}}{\partial\beta_{2}^{n_2}}\beta_{2}^{a_6}\right)\nonumber\\
& \left.
  \left(  \tfrac{\partial^{n_2}}{\partial\beta_{3}^{n_2}}\beta
_{3}^{a_7}\right)
\left(  \tfrac{\partial^{n_2}}{\partial\beta_{4}^{n_2}}
\beta_{4}^{a_8}\right)  \right\vert _{\left(  \alpha
_{i},\beta_{i}\right)  =\left(  0,0\right)  }\tag{A1}
\end{align}
where different parameters are given by
\beqar
a_1 &=&2i_{11}+i_{3}-i_{4}+l_{1}-l_{2}+l-l_{1}+r_{1}-r_{2}+r-r_{1}+j_{3}+k-k_{1}\nonumber\\
a_2&=& 2i_{10}-2i_{11}+i_{2}-i_{3}+l_{7}+l_{6}-l_{7}+r_{6}-r_{7}%
+r_{5}-r_{6}+j_{3}+k_{3}\nonumber\\
a_3&=&2i_{9}-2i_{10}+i_{3}-i_{4}+l_{5}-l_{6}+l_{4}-l_{5}+r_{7}+r_{4}-r_{5}%
+j_{2}-j_{3}+k_{3}\nonumber\\
a_4 &=&i_{2}-i_{3}+2i_{8}-2i_{9}+j_{2}-j_{3}+l_{3}-l_{4}%
+l_{2}-l_{3}+k-k_{1}+r_{3}-r_{4}+r_{2}-r_{3}\nonumber\\
a_5&=& r_{1}-r_{2}+i_{1}-i_{2}+l_{6}-l_{7}+l_{3}-l_{4}+2i_{7}-2i_{8}+r_{4}%
-r_{5}+j_{1}-j_{2}+k_{2}-k_{3}\nonumber\\
a_6&=& l_{5}-l_{6}+i-i_{1}+l_{1}-l_{2}%
+2i_{6}-2i_{7}+r_{5}-r_{6}+r_{3}-r_{4}+j-j_{1}+k_{2}-k_{3}\nonumber\\
a_7&=& l_{2}-l_{3}+l_{7}+i_{1}-i_{2}+r_{7}+r-r_{1}+j-j_{1}+k_{1}-k_{2}%
+2i_{5}-2i_{6}\nonumber\\
a_8&=&l-l_{1}+l_{4}-l_{5}+i-i_{1}+r_{6}-r_{7}+r_{2}-r_{3}+2i_{4}%
-2i_{5}+k_{1}-k_{2}+j_{1}-j_{2}\nonumber
\eeqar
and for the coherence of notations,
$\left( i_{0},j_{0},k_{0},l_{0},r_{0}\right)  \equiv\left(  i,j,k,l,r\right)$
has to be under heard.
Making use of the well-known formula%
\beq
\frac{\partial}{\partial x^{n}}\left.  x^{l}\right\vert _{x=0}=n!\delta_{l,n}\tag{A2}
\eeq
we end up with {the form}
\begin{align}
P_{n_1n_2}  & =\left(\frac{\lam_{1}\lam_{2}}{\pi} n_1!n_2!\right)^2J
{\textstyle\sum\limits_{i,j,k,l,r=0}^{\infty}}
\left(
{\textstyle\prod\limits_{e=1}^{11}}
{\textstyle\sum\limits_{i_{e}=0}^{i_{e-1}}}
\tfrac{1}{\left(  i_{e-1}-i_{e}\right)  !}\right)  \left(
{\textstyle\prod\limits_{e=1}^{3}}
{\textstyle\sum\limits_{j_{e}=0}^{j_{e-1}}}
\tfrac{1}{\left(  j_{e-1}-j_{e}\right)  !}\right)
\left(
{\textstyle\prod\limits_{e=1}^{3}}
{\textstyle\sum\limits_{k_{e}=0}^{k_{e-1}}}
\tfrac{1}{\left(  k_{e-1}-k_{e}\right)  !}\right) \nonumber \\
& \times \left(
{\textstyle\prod\limits_{e=1}^{7}}
{\textstyle\sum\limits_{l_{e}=0}^{l_{e-1}}}
\tfrac{1}{\left(  l_{e-1}-l_{e}\right)  !}\right)
\left(
{\textstyle\prod\limits_{e=1}^{7}}
{\textstyle\sum\limits_{r_{e}=0}^{r_{e-1}}}
\tfrac{1}{\left(  r_{e-1}-r_{e}\right)  !}\right)  \left(  \frac{u}{\rho
}\right)  ^{i}\left(  \frac{2v}{\rho}\right)  ^{j}\left(  \frac{2w}{\rho
}\right)  ^{k}\left(  \frac{2t}{\rho}\right)  ^{l}\left(  \frac{2s}{\rho
}\right)  ^{r}
\nonumber\\
&\times \frac{2^{i-i_{4}}}{i_{11}!j_{3}!k_{3}!l_{7}!r_{7}!}
\ \delta_{b_1,n_1}\delta_{b_2,n_1} \delta_{b_3,n_1}\delta_{b_4,n_1}
\delta_{b_5,n_2}\delta_{b_6,n_2} \delta_{b_7,n_2}\delta_{b_8,n_2}. \tag{A3} 
\end{align}
This shows clearly that a non vanishing purity should satisfy a set
of constraint on different quantum numbers. These are 
\beq
\left\{
\begin{array}
[l]{llllllllllllllll}%
b_1-n_1= 2i_{11}+i_{3}-i_{4}-l_{2}+l-r_{2}+r+j_{3}+k-k_{1}-n_1=0\\
b_2-n_1= 2i_{10}-2i_{11}+i_{2}-i_{3}+l_{6}-r_{7}+r_{5}+j_{3}+k_{3}-n_1=0\\
b_3-n_1= 2i_{9}-2i_{10}+i_{3}-i_{4}-l_{6}+l_{4}+r_{7}+r_{4}-r_{5}+j_{2}-j_{3}%
+k_{3}-n_1=0\\
b_4-n_1= i_{2}-i_{3}+2i_{8}-2i_{9}+j_{2}-j_{3}-l_{4}+l_{2}+k-k_{1}-r_{4}+r_{2}-n_1=0\\
b_5-n_2=l_{5}-l_{6}+i-i_{1}+l_{1}-l_{2}+2i_{6}-2i_{7}+r_{5}-r_{6}+r_{3}-r_{4}%
+j-j_{1}+k_{2}-k_{3}-n_2=0\\
b_6-n_2= l_{2}-l_{3}+l_{7}+i_{1}-i_{2}+r_{7}+r-r_{1}+j-j_{1}+k_{1}-k_{2}+2i_{5}%
-2i_{6}-n_2=0\\
 b_7-n_2= r_{1}-r_{2}+i_{1}-i_{2}+l_{6}-l_{7}+l_{3}-l_{4}+2i_{7}-2i_{8}+r_{4}%
-r_{5}+j_{1}-j_{2}+k_{2}-k_{3}-n_2=0\\
b_8-n_2= l-l_{1}+l_{4}-l_{5}+i-i_{1}+r_{6}-r_{7}+r_{2}-r_{3}+2i_{4}-2i_{5}+k_{1}%
-k_{2}+j_{1}-j_{2}-n_2=0.
\end{array}
\right. \tag{A4}\nonumber
\eeq
We arrange the labels into two sets 
that we refer to them as the principals and secondary ones,
respectively. The so-called secondary
ones disappear upon summation of the 8 constraints and we get
\beq
i+j+k+l+r=2(n_1+n_2)\tag{A5}
\eeq
which is the constraint on the principal labels.
The main result that emerges is that the purity is only depending on two
parameters, such as
\beq
P_{n_1n_2}\left(  \eta,\theta\right)  =\tfrac{2\left(  \frac{2}{\rho}\right)
^{2(n_1+n_2)} (n_1!n_2!)^2}{\sin  \theta  \sqrt{2\cosh (2\eta)+\tan^{2}
\frac{\theta}{2}  +\cot^{2}  \frac{\theta}{2}  }}%
{\textstyle\sum\limits_{i+j+k+l+r=2(n_1+n_2)}}
C_{n_1n_2}\left(  i,j,k,l,r\right)  u^{i}v^{j}w^{k}t^{l}s^{r}\tag{A6}.
\eeq
The most important future of our result is that the function $C_{n_1n_2}\left(  i,j,k,l,r\right)$ can now be
derived exactly for any $n_1$ and $n_2$. This is
\begin{align}
 C_{n_1n_2} & =\left(
{\textstyle\prod\limits_{e=1}^{11}}
{\textstyle\sum\limits_{i_{e}=0}^{i_{e-1}}}
\frac{1}{\left(  i_{e-1}-i_{e}\right)  !}\right)  \left(
{\textstyle\prod\limits_{e=1}^{3}}
{\textstyle\sum\limits_{j_{e}=0}^{j_{e-1}}}
\frac{1}{\left(  j_{e-1}-j_{e}\right)  !}\right)  \left(
{\textstyle\prod\limits_{e=1}^{3}}
{\textstyle\sum\limits_{k_{e}=0}^{k_{e-1}}}
\frac{1}{\left(  k_{e-1}-k_{e}\right)  !}\right)  \left(
{\textstyle\prod\limits_{e=1}^{7}}
{\textstyle\sum\limits_{l_{e}=0}^{l_{e-1}}}
\frac{1}{\left(  l_{e-1}-l_{e}\right)  !}\right)  \nonumber\\
& \times\left(
{\textstyle\prod\limits_{e=1}^{e=7}}
{\textstyle\sum\limits_{r_{e}=0}^{r_{e-1}}}
\frac{1}{\left(  r_{e-1}-r_{e}\right)  !}\right)  \left(  \frac{2^{-i_{4}%
}\left(  -1\right)  ^{i_{2}-i_{8}}\left(  -1\right)  ^{r-r_{1}+r_{3}%
-r_{5}+r_{6}-r_{7}}\left(  -1\right)  ^{l_{1}-l_{3}+l_{4}-l_{5}+l_{6}-l_{7}}%
}{i_{11}!j_{3}!k_{3}!l_{7}!r_{7}!}\right) \tag{A7}.
\end{align}
Using the above constraints, we show that $C_{n_1n_2}$ can be reduced
to the form 
\begin{align}
C_{n_1n_2}   & =\left(
{\textstyle\prod\limits_{e=1}^{4}}
{\textstyle\sum\limits_{i_{e}=0}^{i_{e-1}}}
\right)  \left(
{\textstyle\prod\limits_{e=1}^{3}}
{\textstyle\sum\limits_{j_{e}=0}^{j_{e-1}}}
\right)  \left(
{\textstyle\prod\limits_{e=1}^{3}}
{\textstyle\sum\limits_{k_{e}=0}^{k_{e-1}}}
\right)  \left(
{\textstyle\prod\limits_{e=1}^{7}}
{\textstyle\sum\limits_{l_{e}=0}^{l_{e-1}}}
\frac{1}{\left(  l_{e-1}-l_{e}\right)  !}\right)
\left(
{\textstyle\prod\limits_{e=1}^{7}}
{\textstyle\sum\limits_{r_{e}=0}^{r_{e-1}}}
\frac{1}{\left(  r_{e-1}-r_{e}\right)  !}\right) \nonumber\\
&
{\frac{ 2^{-i_{4}}
(-1)^{l_{1}-l_{3}+l_{4}-l_{5}+l_{6}-l_{7} + r-r_{1}+r_{3}-r_{5}+r_{6}-r_{7}+ i_2-c_1-c_2}}
{{ \left(  i-i_{1}\right)!\left(  i_{1}-i_{2}\right)  !\left(  i_{2}-i_{3}\right)  !
\left(  i_{3}-i_{4}\right)
! l_{7}! r_{7}! c_3!c_4! c_5! c_6!c_7! c_8!c_9! c_{10}! }}}\tag{A8}
\end{align}
where the involved parameters are fixed as
\beqar
c_1 &=& \frac{1}{2} \left[ 2n_1-\left(  i_{2}%
-i_{3}\right)  -\left(  j_{2}-j_{3}\right)  -\left(  k-k_{1}\right)  -\left(
l_{2}-l_{3}\right)  -\left(  l_{3}-l_{4}\right)  -\left(  r_{2}-r_{3}\right)
-\left(  r_{3}-r_{4}\right)\right.  \nonumber\\
 && \left. 
 \left(  i_{3}-i_{4}\right)  -\left(
r_{4}-r_{5}\right)  -\left(  j_{2}-j_{3}\right)  -\left(  l_{4}-l_{5}\right)
-\left(  l_{5}-l_{6}\right)  -r_{7}-k_{3} \right]\nonumber\\
c_2 &=& \frac{1}{2} \left[ n_1-\left(  i_{2}-i_{3}\right)
-\left(  r_{5}-r_{6}\right)  -\left(  r_{6}-r_{7}\right)  -l_{6}-j_{3}-k_{3}%
 +\left(  i_{3}-i_{4}\right)  -\left(  l-l_{1}\right) \right. \nonumber \\
 && \left. 
 \left(
l_{1}-l_{2}\right)  -\left(  r-r_{1}\right)  -\left(  r_{1}-r_{2}\right)
-j_{3}-\left(  k-k_{1}\right)  \right]\nonumber\\
c_3&=& \left(  \frac{n_1-\left(  i_{3}%
-i_{4}\right)  -\left(  r_{4}-r_{5}\right)  -\left(  j_{2}-j_{3}\right)
-\left(  l_{4}-l_{5}\right)  -\left(  l_{5}-l_{6}\right)  -r_{7}-k_{3}}%
{2}\right)  !\nonumber\\
c_4 &=& \left(  \frac{n_1-\left(  i_{2}-i_{3}\right)  -\left(  r_{5}%
-r_{6}\right)  -\left(  r_{6}-r_{7}\right)  -l_{6}-j_{3}-k_{3} } {2}\right)!\nonumber\\
 c_5 &=& \left(\frac{n_1-\left(  i_{3}-i_{4}\right)  -\left(
l-l_{1}\right)  -\left(  l_{1}-l_{2}\right)  -\left(  r-r_{1}\right)  -\left(
r_{1}-r_{2}\right)  -j_{3}-\left(  k-k_{1}\right)  }{2}\right)  !\nonumber\\
c_6 &=& \left(
\frac{n_1-\left(  i_{2}-i_{3}\right)  -\left(  j_{2}-j_{3}\right)  -\left(
k-k_{1}\right)  -\left(  l_{2}-l_{3}\right)  -\left(  l_{3}-l_{4}\right)
-\left(  r_{2}-r_{3}\right)  -\left(  r_{3}-r_{4}\right)  }{2}\right)  !\nonumber\\
c_7 &=& \left(  \frac{n_2-\left(  r_{1}-r_{2}\right)  -\left(
i_{1}-i_{2}\right)  -\left(  l_{6}-l_{7}\right)  -\left(  l_{3}-l_{4}\right)
-\left(  r_{4}-r_{5}\right)  -\left(  j_{1}-j_{2}\right)  -\left(  k_{2}%
-k_{3}\right)  }{2}\right)  !\nonumber\\
c_8 &=& \left(  \frac{n_2-\left(  l_{5}-l_{6}\right)
-\left(  i-i_{1}\right)  -\left(  l_{1}-l_{2}\right)  -\left(  r_{5}%
-r_{6}\right)  -\left(  r_{3}-r_{4}\right)  -\left(  j-j_{1}\right)  -\left(
k_{2}-k_{3}\right)  }{2}\right)  !\nonumber\\
c_9 &=& \left(  \frac{n_2-\left(  l-l_{1}\right)  -\left(  l_{4}%
-l_{5}\right)  -\left(  i-i_{1}\right)  -\left(  r_{6}-r_{7}\right)  -\left(
r_{2}-r_{3}\right)  -\left(  k_{1}-k_{2}\right)  -\left(  j_{1}-j_{2}\right)
}{2}\right)  !\nonumber\\
c_{10} &=& \left(  \frac{n_2-\left(  l_{2}-l_{3}\right)  -\left(  i_{1}%
-i_{2}\right)  -\left(  r-r_{1}\right)  -\left(  j-j_{1}\right)  -\left(
k_{1}-k_{2}\right)  -r_{7}-l_{7} }{2}\right)!\nonumber
\eeqar


\end{document}